\documentclass[a4paper,12pt]{article}

\usepackage[latin2]{inputenc}
\usepackage{graphicx}
\usepackage{amstext}
\usepackage[]{ntheorem}
\usepackage{fancyhdr}
\usepackage{amsfonts,dsfont}
\usepackage{amssymb}
\usepackage{amsmath,chemarrow}
\usepackage{array}
\usepackage{multirow}

\usepackage{t1enc}
\usepackage{tikz}
\usepackage{hyperref}
\usepackage[affil-it]{authblk}

\newcommand{\RR}{\mathds R}
\newcommand{\BB}{\mathcal{B}}

\title{Introduction of flexible production bids and combined package-price bids in a framework of integrated power-reserve market coupling}

\author{D\'{a}vid Csercsik}

\affil{P\'{a}zm\'{a}ny P\'{e}ter Catholic University\\ Faculty of Information Technology and Bionics\\ Pr\'{a}ter~u. 50/A 1083 Budapest, Hungary \\
              Tel.: +36-1 886 47 00
              Fax: +36-1 886 47 24\\
              \emph{csercsik@itk.ppke.hu}}

\begin{document}

\maketitle

\abstract{In this article we propose a multi-zonal integrated energy-reserve market model. We assume that bidders may submit their demand and supply bids on the one hand in the form of conventional hourly step bids and block bids, which are cleared and paid according to market clearing prices (MCPs). On the other hand, suppliers may submit so called flexible production bids, while both suppliers and consumers may submit fill-or-kill type package-priced combined bids -- these bids are accepted if their acceptance implies an improvement in the resulting total social welfare, which the market clearing algorithm aims to optimize. The model includes network constraints for the nominal case (if no reserves are activated) and also for perturbed cases when the allocated reserves are activated.}


\section{Introduction}
\label{Introduction}

The most important aim of electricity markets is to harmonize power demand and supply in a way, which in ideal case results in the highest possible social welfare (SW).
The concept of SW basically originates from the 'pay as clear' principle \cite{Son2004ShortTerm}.
In the most simple framework for electricity market clearing, supply and demand bids are submitted for
a single period of the trading interval, each bid being described by two parameters, the bid quantity and the bid price per unit (PPU). In the case of demand bids, the product of the bid quantity and the PPU describes the willingness to pay for the required quantity, while in the case of supply bids, the same product corresponds to the minimal required income for the offered amount (usually assumed to be equal to the cost of the production). The market is cleared according to the so called \emph{market clearing price} (MCP):
Demand bids with bid PPU lower than the MCP will be rejected, as well as supply bids with bid PPU higher than the MCP. Bids whose PPU is equal to the MCP may be also partially accepted.

\begin{figure}[h!]
  \centering
  \includegraphics[width=7cm]{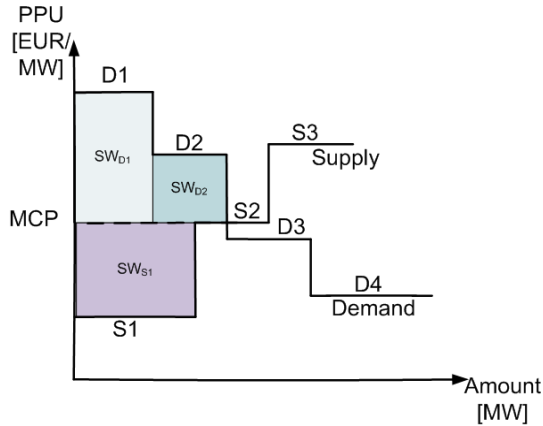}\\
  \caption{Social welfare of single bids in the one-period market model. S$i$ and D$i$ correspond to supply and demand bids, while MCP stands for the market clearing price.}\label{basic_SW}
\end{figure}

Bids are paid for according to the MCP, which means that the bidder, e.g. in the case of an accepted demand bid, pays less for the required quantity compared to his willingness to pay (and similarly, supply bidders receive potentially more payment for accepted supply bids). This surplus, which is the product of the difference between the bid PPU and the MCP and the bid quantity, is called the social welfare (SW) of the bid \cite{madani2017revisiting,madani2014minimizing}. The total social welfare (TSW) of a dispatch is the sum of SW values corresponding to single bids, and may be represented as the area between the supply and the demand curve as depicted in Fig. \ref{basic_SW}.

In the simple one-period example depicted in Fig. \ref{basic_SW}, the maximization of TSW is trivial: We determine the MCP from the intersection of supply and demand curves (this will ensure the energy balance), and clear the market according to this MCP. There are two factors which make the problem more complex. On the one hand, the clearing is performed simultaneously for multiple periods, and there are bids which imply interconnections between different periods (e.g. they have the 'fill or kill' property -- they must be accepted in all included periods or must be completely rejected). Minimum income condition orders and scheduled stop conditions can imply similar interconnections between periods \cite{sleisz2019new}.
 On the other hand, we may have multiple price zones, connected with transmission lines. In this case the energy balance is not required for every single price zone, but it must hold for the total system, while the transmission constraints of the connecting lines must be taken into account \cite{motto2002network}.


In addition, operators of the power system have to ensure the stability and security. In the current setup, as we assume the central authority operates the market with regard to the transmission system as well, we will use the terminology of independent system operator (ISO).

Stability refers to frequency stability \cite{zhao2014design} or voltage stability \cite{VanCutsem1998}, while security refers to e.g. n-1 line and node contingency, which means that if one of the lines or one of the nodes of the network fails instantly, the resulting flows may not overload any of the remaining lines \cite{khanabadi2013optimal}.
The stability of frequency is dependent on the supply-demand balance: If consumers or suppliers deviate from their predefined schedule, the ISO activates previously allocated (positive or negative) reserves at generating units to restore the balance.


These reserves practically mean rights for the ISO to give orders to generating units to increase or decrease actual generation values.
In most of the countries where a liberalized electricity trade takes place, separate markets were created for the allocation of such and other reserves, called altogether ancillary services \cite{raineri2006technical}.

Joint (or integrated) energy and reserve markets are representing a concept, where the allocation of power and reserve to generating units takes place not on disjoint markets, but in one integrated auction \cite{gonzalez2014joint}.

One main benefit of integrated markets is described in \cite{Sores2014Day-ahead} as:  '\emph{co-optimization  enables  the participants  to  achieve  more  surplus  by  providing  an  efficient
way  to  submit  all  possible  combinations  of  energy-reserve allocation  to  the  market.  Therefore  the  risk  of  precommitting generating  capacity  to  sequential  offers  of  different  products
and  clearing  can  be  eliminated}'.
The paper \cite{galiana2005scheduling} formulates a similar consideration as '\emph{Since distinct reserve
services can in fact be strongly coupled, and the heuristics required to bridge the various sequential markets can ultimately lead to loss of social welfare, simultaneous energy/reserves market-clearing
procedures have been proposed and are in use. However, they generally schedule reserve services subject to exogenous rules and parameters that do not relate to actual operating conditions}'.

While several results have been already published in the field of integrated markets, the presented approaches usually are driven by the unit-commitment spirit of North American market models,
where the generating units are not self-scheduling. A not self-scheduling clearing means that generating units submit technical characteristics and production costs to the ISO who determines production levels and reserve allocations according to these parameters.

The main aim of the current paper is to provide a possible framework for multi-node integrated markets, but in contrast to the cost minimization approach used e.g. in \cite{wu2004pricing}, we aim at maximizing the total social welfare (SW).
The paper \cite{galiana2005scheduling} proposes a security constrained simultaneous clearing of energy and reserve services with a perturbation approach similar to the one proposed in the current paper. This paper considers primary, secondary and tertiary reserves as well and uses generals function for the description of the social welfare. The supply side in \cite{galiana2005scheduling} is also formulated in a unit-commitment spirit. The approach of the paper \cite{arroyo2005energy} is similar, it also proposes that at any given network bus all scheduled reserve types should be priced not at separate rates but at a
common rate equal to the marginal cost of security at that bus. The paper \cite{amjady2009stochastic} uses a multiobjective mathematical programming (MMP) approach including MCPs as well in the formulation. The paper \cite{aghaei2009multi} uses also an MMP approach, defining the MCP only for the energy.

In contrast, in European type markets, the self-scheduling generating units may bid with a variety of products, and act like more active market participants \cite{biskas2014european}. An approach for co-optimizing power and reserve allocation which is motivated by this type of power market is described in the articles \cite{Sores2014Day-ahead,polgari2015new,divenyi2019algorithm}.

In the day-ahead market, where the clearing is determined for 24 consecutive hours, technological considerations of generating units imply further challenges (in integrated and conventional markets as well).
Startup costs and minimal operating loads are the most common sources of non-convexities, but we may also think of minimal up and down times. These non-convexities are usually handled by the introduction of block orders,
which may be rejected or accepted in a binary manner (no partial acceptance is allowed), thus the representative variables in the clearing are binary.

An approach to represent the constant and variable costs (corresponding to start-up and production
respectively) of generating units is the concept of minimum income condition (MIC) orders \cite{Contreras2001,sleisz2014clearing,sleisz2015efficient,sleisz2015challenges,sleisz2017integrated}. As described in \cite{sleisz2015challenges}, '\emph{minimum income orders are supply orders consisting of several hourly step bids for potentially different market hours, and they are bound together by the MIC which prescribes that the overall income of the MIC order must cover its given costs}'. The efficient clearing of such bids is described in \cite{sleisz2017integrated}. In this framework, generation costs corresponding to this type of bid are zero if the bid is rejected, otherwise  they  are considered  with a  fixed and a linear  variable term which are determined by the bidder. Incomes in the case of the proposed MIC bid can be expressed as the product of accepted quantities and MCPs. In this concept, since the elements of the MIC bids are standard hourly step bids, the generation profile of the unit submitting the MIC bid is fully determined by the MCPs.


In this paper we propose a somewhat different approach, namely we introduce the concept of \emph{flexible production bids} (shortly FP bids) and combined bids.
As we will see, flexible production bids are formed in the spirit of unit commitment: The
production values for the single periods are determined by the ISO during the clearing, considering the technical and cost parameters of the unit. Technical parameters are the load gradient constraints, while the cost parameters are the start-up and variable cost values.
Combined bids in contrast hold fixed quantities of power and reserve and are cleared and paid as a whole package if accepted.

In addition we also consider the coupling of combined power-reserve markets, which mean that we formulate transmission constraints on nominal flows and also flows originating from the activation of reserves.
We use a linear DC load flow based framework to formulate the transmission constraints.

The proposed framework may be also considered as a kind of transition between European and US type markets in the sense that on the one hand conventional price-quantity (step) bids are submitted, and on the other hand generating units may also submit generation characteristics in the form of FP bids, in which case their power and reserve allocation will be scheduled by the ISO. In addition, participants may also submit fixed-price combined bids, which represent basically pay-as-bid type bids.

Since the problem formulation in itself is a complex challenge (even if we would consider only 'conventional' coupling of integrated markets without innovative bid types), in this paper we confine ourselves to present only the details of the formulation, and only shortly discuss the computational properties and demands of the resulting optimization problem.

\section{The market model}
\label{MM}
The notations used through the paper are summarized in Appendix A.
The basis of the proposed framework is a standard uniform price (European type) multi-node (or in other words zonal) electricity market model with $T$ time periods (see e.g. the basic structure in \cite{madani2017revisiting}).

  In the current paper we only consider reserves corresponding to frequency control. Furthermore, to keep the initial model formulation tractable we focus only on secondary reserves.
  We assume that reserve-providing units are paid for the allocation of reserve capacities, in other words in the current model we do not take into account if reserves are activated or not.
  The possibility of allocating secondary and tertiary reserves simultaneously in the proposed framework is discussed in section \ref{Discussion}.

 \subsection{Bid types in the model}
We suppose in the following that one period of the model corresponds to one hour.

 \subsubsection{One hour bids}

\paragraph{One-hour single-product bids} These bids are the principal elements of the market model. They describe demand or supply of a single product (power, positive or negative reserve) in a single time period, and their acceptance is independent of the acceptance of bids regarding other time periods.

It can be assumed that most bids are submitted in this format to the market. These bids are characterized by a quantity ($B$), by the index of the time period in which the bid is relevant ($t$) and a respective price (per unit), denoted by $\Theta$.

Such one-hour bids are cleared according to MCPs denoted by $\varphi_i^{P}(t)$, $\varphi_i^{Rp}(t)$ and $\varphi_i^{Rn}(t)$ corresponding to power, positive and negative reserve respectively in each node $i$, regarding the respective time period $t$.
If the resulting MCP is equal PPU of a bid, the partial acceptance of the bid is allowed, formally the bid acceptance indicator $y$  is $\in [0,1]$ in this case.

\subsubsection{Multiple period bids}
\label{Multiple period bids}

Under multiple period bids we mean bids which may include multiple periods as well, but must be taken into account and cleared as a single bid.

 \paragraph{Block bids} In our terminology, under block bids  we mean single product (power or reserve) bids, which include multiple (consecutive) time periods. We assume that block bids have the fill-or-kill property, in the sense that either the total offered quantity is either fully accepted for all respective time periods, or the bid is completely rejected.

These bids are characterized by quantities for the corresponding hours ($B$ --  a vector in this case), by the indices of the time periods in which the bid is relevant ($t$) and the respective PPUs, denoted by $\Theta$ (also a vector). Although in the practice the bid quantities and PPUs usually are the same for every period of the bid, the vector formulation allows potentially different quantities and PPUs for each period.

The acceptance constraint in the case of block bids is that the resulting total SW must be positive \cite{madani2014minimizing}. The total SW of a block bid is the sum of the SWs corresponding to the included time periods.  Block bids are very common in electricity markets and they are discussed e.g. in \cite{meeus2009block,madani2014minimizing}.

\paragraph{Remark: Standard bids}
In the following, under standard bids we mean to 1-hour single-product bids or block bids. We distinguish these bids from the bid types described in the following, since their acceptance is explicitly determined by MCPs.

\paragraph{Flexible production bids} Flexible production or FP bids are suited for generating units who practically offer their generating capacity in a unit-commitment type offer. Upon the acceptance of such bids, the ISO assigns nonzero power and reserve amounts to units submitting these bids for each period included in the bid, according to the actual needs of the market.
These bids are characterized in the proposed model by start-up cost ($\alpha$), variable cost ($\beta$) and ramp constraints ($RU$ for ramp-up and $RD$ for ramp-down). The maximal possible amount of assigned reserve is determined by the assigned power production profile and by the ramp constraints (e.g. if in two consecutive periods the output of the unit according to the power production profile is increased with $RU$, no positive reserve may be assigned to it).

If an FP bid is accepted, the generating unit is paid off according to produced quantities (determined by the ISO) and respective MCPs, and its income must cover the reported expenses of generation, derived from start-up and variable costs. The formulation of last consideration may be viewed as a variant of the so called minimum income condition (MIC)
\cite{sleisz2014clearing,sleisz2015efficient,sleisz2015challenges}.

\paragraph{Example 1} To illustrate the concept of FP bids, let us consider a simple 2 period example.
Let us assume that the set of standard (in this case only 1-hour) bids is as summarized in Table
\ref{table_FP_example}.

\begin{table}[h!]
\begin{center}
\begin{tabular}{|c|c|c|c|}
  \hline
  bid ID & relevant period & quantity (B) & PPU ($\Theta$) \\ \hline
  $D^1_1$ & 1 & 15 & 90 \\
  $D^1_2$ & 1 & 20 & 80 \\
  $S^1_1$ & 1 & 27 & 75 \\
  $S^1_2$ & 1 & 13 & 85 \\
  $D^2_1$ & 2 & 15 & 90 \\
  $D^2_2$ & 2 & 20 & 80 \\
  $S^2_1$ & 2 & 27 & 75 \\
  $S^2_2$ & 2 & 13 & 85 \\
  \hline
\end{tabular}
\caption{Standard bids in of example 1 (the upper index in the bid ID refers to the period)\label{table_FP_example}}
\end{center}
\end{table}

\begin{figure}[h!]
  \centering
  \includegraphics[width=7cm]{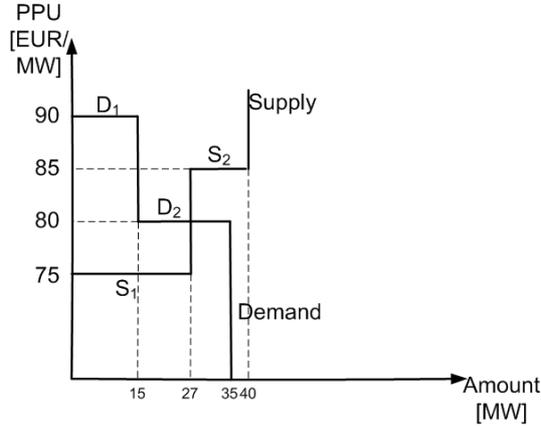}\\
  \caption{Supply-demand curves of example 1}\label{FP_bid_1}
\end{figure}

We can see in Table \ref{table_FP_example} that the standard bids are the same for hour 1 and 2, thus they imply the supply-demand curves depicted in Fig. \ref{FP_bid_1} for both hours.

First let us assume the scenario where no other bids are present. In this case, the dispatch calculation is very simple: The MCP (denoted by $\varphi$) is determined by the intersection of the curves ($\varphi$=80): $D_1$ and $S_1$ will be fully accepted while $D_2$ will be partially accepted.
The social welfare of the demand and supply side in each period may be calculated as:

\begin{align*}
&SW^{D}=(90-80)15=150\\
&SW^{S}=(80-75)27=135~~~~,
\end{align*}
thus the total social welfare equals to 285 for each period thus $TSW$=570.

On the other hand, let us assume a scenario, when in addition to the standard bids,
an FP bid is also present with the parameters $\alpha=3000$, $\beta=28$ (we can assume that $RU$ and $RD$ are arbitrary positive values).

In this case let us consider the scenario where $\varphi=72$ for both periods. According to the MCP, the standard supply bids will be rejected (as their PPU is higher than the MCP), while both demand bids will be fully accepted, resulting in the social welfare
\begin{align*}
SW^{D}=(90-72)15 + (80-72)20=430
\end{align*}
for both periods.
Regarding the supply side, the unit corresponding to the FP bid has to produce 35 MW in each period. The cost of the production of the FP bid may be calculated as
$ \alpha + 2 \beta 35=4960 $, while the income of the FP bid is $35 \cdot 72 \cdot 2=5040$.
The income of the bid covers the production cost (which is a necessary condition for the acceptance of the FP bid), and $SW^S=80$. In this case $TSW=860+80=940$.

As the $TSW$ is higher in the case of the second scenario (940 vs 570), if the FP bid is also present in the market, the market clearing algorithm will prefer the second solution, as it aims to maximize the $TSW$. In general, in order to maximize the TSW, the market clearing algorithm has to determine the MCPs and the acceptance of FP bids simultaneously.

 Let us note furthermore that the acceptance of the FP bid is not explicitly determined by the MCP:
 If we lower the PPU of the first and the second supply bid to 60 and 72 respectively, and suppose $\varphi=72$, both demand bids will be accepted (while $S_2$ will be partially accepted to ensure the power balance) and we get

\begin{align*}
&SW^{D}=(90-72)15+(80-72)20=430\\
&SW^{S}=(72-60)27+=324
\end{align*}
for each period, resulting in $TSW=1508$, a solution clearly preferable compared to the acceptance of the FP bid.

Regarding the notations corresponding to FP bids in the model, as in general we consider multiple nodes, we assume that each FP bid corresponds to a generating unit in a certain node of the network.
Each node may hold multiple generating units, but not all nodes necessarily hold generating units.
We use binary variables to describe whether a generating unit operates or not in a given time period.
 $v_{ij}(t)$ denotes the activity indicator of unit $j$ of node $i$ at time $t$.
 Units are indexed from 1 in each node. For example if there are 2 units in node 1 and 1 unit in node two,
 we will have variables $v_{11}(t)$, $v_{12}(t)$ and $v_{21}(t)$ for each time period $t$.
 With the help of these binary variables we can describe start-up costs, minimal up and down times and minimal load of units. On the other hand, we suppose that if the corresponding activity indicator is 0, the output of the unit is 0 (regarding power, and both types of reserves as well). $P_{ij}(t)$ denotes the power production value allocated to unit $j$ of node $i$ at time $t$.
Regarding reserves, $Rp_{ij}(t)$ and $Rn_{ij}(t)$ denote respectively the positive and negative reserve value allocated to unit $j$ of node $i$ at time $t$.

\paragraph{Combined bids} Combined bids in the proposed framework make possible to submit bids simultaneously for power and reserve production (or consumption).
In the case of combined bids, the bid holds fixed values of power, positive and negative reserves, potentially including multiple time periods. The parameters of this bid type are the amounts of products offered for the respective time periods, and total price. The price is not interpreted as per unit in this case, but as a total amount, which shall be at least paid to the bidder upon acceptance -- independent of MCPs. In addition to this fixed price, to each combined bid a nonnegative surplus is assigned by the MO (see the details later).

We assume that combined bids have the fill-or-kill property and we call the standard bids and combined bids \emph{fixed quantity} (FQ) bids (in contrast to FP bids where the quantity is assigned to the bid by the ISO).

\paragraph{Example 2}
To illustrate the concept of combined bids, let us suppose a single-period scenario, where
the standard power and reserve bids are as summarized in table \ref{table_C_example} (the power bids define the same supply-demand curves as in Example 1). In the case of this simple example we consider only one type of reserve (arbitrarily + or -).

\begin{table}[h!]
\begin{center}
\begin{tabular}{|c|c|c|}
  \hline
  bid ID& quantity (B) & PPU ($\Theta$) \\ \hline
  $D^P_1$ & 15 & 90 \\
  $D^P_2$ & 20 & 80 \\
  $S^P_1$ & 27 & 75 \\
  $S^P_2$ & 13 & 85 \\
  $D^R_1$ & 10 & 50 \\
  $D^R_2$ & 10 & 40 \\
  $S^R_1$ & 15 & 45 \\
  \hline
\end{tabular}
\caption{Standard bids in of example 2 (the upper index refers to power/reserve)\label{table_C_example}}
\end{center}
\end{table}

Again, let us first assume the scenario where no other bids are present.
In this case $\varphi^P=80$, $\varphi^R=45$, resulting in $SW^P=285$ and $SW^R=50$ ($TSW=335$) -- the balance is 27 MW regarding power and 10 MW regarding the reserve.

On the other hand, if in addition to the standard bids we also assume a combined bid offering 15 MW of power and 15 MW of reserve at the price of 1600, the following dispatch is possible.
Regarding the power balance, if $\varphi^P=75$, both demand bids are accepted resulting in the demand of 35 MW, from which 20 MW of power is supplied from the first standard supply bid (which is partially accepted), and the rest from the combined bid.

Regarding the reserve balance, the standard reserve supply bid is rejected, the first standard reserve demand bid is fully accepted while the second one is partially accepted. All 15 MWs of reserve are supplied by the accepted combined bid.

Here we have to check two conditions. First the total income from demand bids must cover the
total cost of supply. The income from power demand bids is $(15+20)75=2625$, while the income from reserve demand bids is $(10+5)40=600$, thus the total income is $3225$.
The cost of the standard power supply bid is $75 \cdot 20=1500$, while the cost of the combined bid is 1600 The total cost is 3100 -- the difference between the total income and the total cost (125) will be assigned to the surplus of the combined bid in this case.

Second, the $TSW$ must exceed the $TSW$ of the first scenario in order to make the dispatch more desirable for the clearing algorithm.

\begin{align*}
&SW^{P}=(90-75)15 + (80-75)*20=325\\
&SW^{R}=(50-40)10 =100~~~~,
\end{align*}
while the SW of the combined bid is equal to its surplus (125), thus $TSW=540>335$.

\subsubsection{Overview of bids}

\paragraph{Fixed quantity and flexible production}
Except for FP bids, for all bids we can say that we know how much they will contribute to power and reserve balances upon their acceptance (partial acceptance is allowed only in the case of one-hour single-product bids). We can call these bids fixed quantity (FQ) bids. We assume that demand bids are always FQ.
The set $\BB$ collets all FQ bid types, regarding the traded product (not distinguishing between one-hour and multiple hour bids).
\begin{align}
& \BB = \{DP,~SP,~DRp,~SRp,~DRn,~SRn,~DC,~SC\}
\end{align}
The first letter stands for demand or supply, while the rest stand for power (P), positive reserve (Rp),
negative reserve (Rn) or combined bids (C). These abbreviations are used through the paper.

\begin{figure}[h!]
  \centering
  \includegraphics[width=7cm]{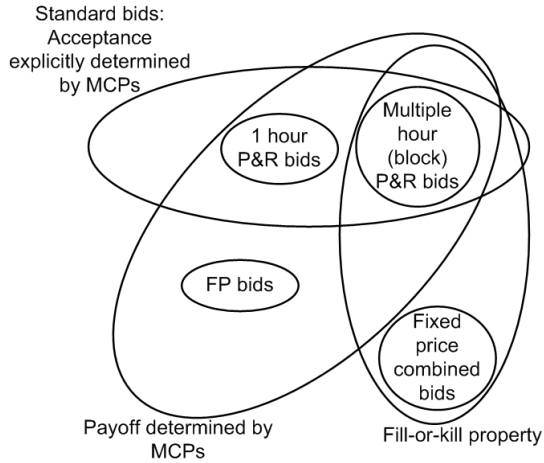}\\
  \caption{Bid types in the proposed market formulation. P\&R stands for power and reserve.}\label{bid_types}
\end{figure}

Figure \ref{bid_types} summarizes the bid types used through the paper and their properties.

\subsection{Clearing of the market}

\begin{figure}[h!]
  \centering
  \includegraphics[width=7cm]{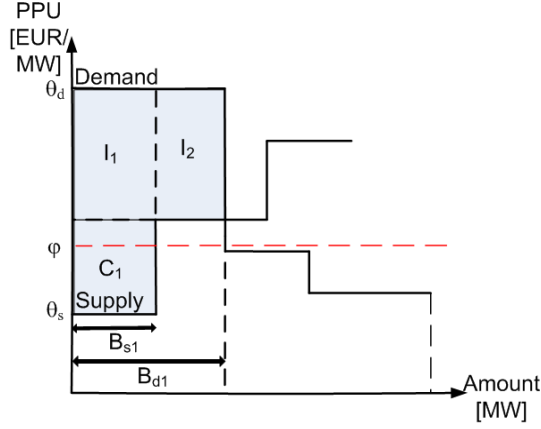}\\
  \caption{Power and income/cost imbalances caused by the particular depicted MCP. $\varphi$ stands for the MCP, while $\theta_d$ and $\theta_s$ denotes the demand and supply bid prices.}\label{spotM_1}
\end{figure}

We may depict the one-hour single product (e.g. power) demand and supply bids for any particular hour in the standard spot-market fashion like in Fig. \ref{spotM_1}. By such an ordering of bids (increasing by PPU in the case of supply and decreasing in the case of demand), if there are enough bids for the curves to intersect in every hour, setting the MCPs according to the intersection prices clears the market (in this case however no block bids, FP bids or combined bids are taken into account, thus all of such bids are rejected).

On the other hand if we consider a scenario of the MCP depicted in Fig. \ref{spotM_1}, we can see that there is an imbalance both in the supplied/consumed power ($B_{d1}-B_{s1}$), and regarding incomes/costs as well. The total income is $I_1+I_2=B_{d1}\varphi$ while the total cost of the accepted supply bid is
$C_1=B_{s1}\varphi$. To put the principle of the clearing very short, we may use the excess income (summed for all hours) to pay for block, FP and combined bids which cover the hourly power/reserve imbalances (as detailed in Example 1 of subsection \ref{Multiple period bids} and depicted in Fig. \ref{FP_bid_1})

The task of the market-clearing algorithm is to find such MCPs, and such scheduling of FP, block and combined bids (via determination of their scheduling/acceptance variables\footnote{To be more precise, in the case of combined bids, the payoff variables have to be determined as well}), which maximizes the total social welfare, and respects the hourly power and reserve balance constraints as well as the network bottlenecks.


\subsection{Assumptions regarding the multi-nodal market structure}
The following assumptions determine the size of the variable vectors.
\begin{itemize}
\item We assume $N$ nodes.
\item The number of units at node $i$ is denoted by $n_i$. We assume that each unit submits a FP bid. The total number of units, which do submit FP bids is denoted by $n$.

\begin{align}
& \sum_i n_i=n
\end{align}

\item $m^{DP}_i$ denotes the number of standard demand power bids at node $i$.
\item $m^{SP}_i$ denotes the number of standard supply power bids at node $i$.

\item $m^{DRp}_i$ denotes the number of standard positive reserve demand bids at node $i$.
\item $m^{SRp}_i$ denotes the number of standard positive reserve supply bids at node $i$.

\item $m^{DRn}_i$ denotes the number of standard negative reserve demand bids at node $i$.
\item $m^{SRn}_i$ denotes the number of standard negative reserve supply bids at node $i$.

\item $m^{DC}_i$ denotes the number of combined demand bids at node $i$.
\item $m^{SC}_i$ denotes the number of combined supply bids at node $i$.

\item $m$ denotes the total number of $FQ$ bids.

\end{itemize}

Furthermore we define the following variables.
\begin{small}
\begin{align}
& m^{DP} = \sum_i m^{DP}_i ,~~~m^{SP} = \sum_i m^{SP}_i  \nonumber \\
&  m^{DRp} = \sum_i m^{DRp}_i,~~~m^{SRp} = \sum_i m^{SRp}_i,~~~m^{DRn} = \sum_i m^{DRn}_i,~~~m^{SRn} = \sum_i m^{SRn}_i  \nonumber \\
&  m^{DC} = \sum_i m^{DC}_i,~~~m^{SC} = \sum_i m^{SC}_i \nonumber \\
& m=m^{DP}+m^{SP}+m^{DRp} +m^{DSp} +m^{DRn} m^{SRn} + m^{DC} + m^{SC}
\end{align}
\end{small}
\subsection{Variables of the model}

In this subsection, according to the previous considerations, we enumerate the variables of the proposed framework.

\subsubsection{Variables corresponding to the clearing of FP bids}

\begin{itemize}

\item $v_{ij}(t)$ denotes the up indicator of unit $j$ of node $i$ at time $t$, equals to 1 if unit $j$ of node $i$ is operating at time period $T$ and zero otherwise. The vector of all such variables is denoted by $v \in \{ 0,1\}^{n (T+1)}$. $v_{ij}(O)$ is an auxiliary variable, which is equal to one if the unit is up in any of the periods in the analyzed time frame (used for the calculation of start-up costs). The structure of $v$ is as follows.
\begin{small}
\begin{align}
v=\left(
    \begin{array}{c}
      v_1 \\
      v_2 \\
      \vdots \\
      v_N \\
    \end{array}
  \right)
  ~~~
  v_i=\left(
    \begin{array}{c}
      v_{i1} \\
      v_{i2} \\
      \vdots \\
      v_{in_i} \\
    \end{array}
  \right)
  ~~~
  v_{ij}=\left(
    \begin{array}{c}
      v_{ij}(1) \\
      v_{ij}(2) \\
      \vdots \\
      v_{ij}(T) \\
      v_{ij}(O)
    \end{array}
  \right)
\end{align}
\end{small}

\item $P_{ij}(t)$ denotes the power production value allocated to unit $j$ of node $i$ at time $t$,
$Rp_{ij}(t)$ denotes the positive reserve allocated to unit $j$ of node $i$ at time $t$, and
$Rn_{ij}(t)$ denotes the negative reserve allocated to unit $j$ of node $i$ at time $t$.
$P \in \{ 0,1\}^{n T}$, $Rp \in \{ 0,1\}^{n T}$, $Rn \in \{ 0,1\}^{n T}$.

\begin{small}
\begin{align}
P=\left(
    \begin{array}{c}
      P_1 \\
      P_2 \\
      \vdots \\
      P_N \\
    \end{array}
  \right)
  ~~~
  P_i=\left(
    \begin{array}{c}
      P_{i1} \\
      P_{i2} \\
      \vdots \\
      P_{in_i} \\
    \end{array}
  \right)
  ~~~
  P_{ij}=\left(
    \begin{array}{c}
      P_{ij}(1) \\
      P_{ij}(2) \\
      \vdots \\
      P_{ij}(T)
    \end{array}
  \right)
\end{align}

\begin{align}
Rp=\left(
    \begin{array}{c}
      Rp_1 \\
      Rp_2 \\
      \vdots \\
      Rp_N \\
    \end{array}
  \right)
  ~~~
  Rp_i=\left(
    \begin{array}{c}
      Rp_{i1} \\
      Rp_{i2} \\
      \vdots \\
      Rp_{in_i} \\
    \end{array}
  \right)
  ~~~
  Rp_{ij}=\left(
    \begin{array}{c}
      Rp_{ij}(1) \\
      Rp_{ij}(2) \\
      \vdots \\
      Rp_{ij}(T)
    \end{array}
  \right)
\end{align}

\begin{align}
Rn=\left(
    \begin{array}{c}
      Rn_1 \\
      Rn_2 \\
      \vdots \\
      Rn_N \\
    \end{array}
  \right)
  ~~~
  Rn_i=\left(
    \begin{array}{c}
      Rn_{i1} \\
      Rn_{i2} \\
      \vdots \\
      Rn_{in_i} \\
    \end{array}
  \right)
  ~~~
  Rn_{ij}=\left(
    \begin{array}{c}
      Rn_{ij}(1) \\
      Rn_{ij}(2) \\
      \vdots \\
      Rn_{ij}(T)
    \end{array}
  \right)
\end{align}
\end{small}

\subsubsection{Variables corresponding to MCPs and bid acceptance indicators}

\item $\varphi_i^P(t)$ denotes the MCP of power at node $i$ at time $t$.

\item $\varphi_i^{Rp}(t)$ denotes the MCP of positive reserve at node $i$ at time $t$.

\item $\varphi_i^{Rn}(t)$ denotes the MCP of negative reserve at node $i$ at time $t$.

\begin{small}
\begin{align}
\varphi=\left(
    \begin{array}{c}
      \varphi^P \\
      \varphi^{Rp} \\
      \varphi^{Rn}
    \end{array}
  \right) ~~~~
  \varphi^P=\left(
    \begin{array}{c}
      \varphi^P_1 \\
      \varphi^P_2 \\
      \vdots \\
      \varphi^P_N
    \end{array}
  \right)
  \varphi^P_i=\left(
    \begin{array}{c}
      \varphi^P_i(1) \\
      \varphi^P_i(2) \\
      \vdots \\
      \varphi^P_i(T)
    \end{array}
  \right)
\end{align}
\end{small}
$\varphi \in \RR_+^{3NT}$. $\varphi^{Rp}$ and $\varphi^{Rn}$ are similarly derived.

\item $y^b_{ij}$ is the bid acceptance indicator of the standard bid of type $b \in \BB$, corresponding to $j$-th such bid of node $i$. $y^b_{ij} \in \{0,1\}$ in the case of fill-or-kill bids.

\begin{small}
\begin{align}
y=\left(
    \begin{array}{c}
      y^{DP} \\
      y^{SP} \\
      y^{DRp} \\
      y^{DRn} \\
      y^{SRp} \\
      y^{SRn} \\
      y^{DC} \\
      y^{SC} \\
    \end{array}
  \right)
  ~~~
y^{DP}=\left(
    \begin{array}{c}
    y^{DP}_1 \\
    y^{DP}_2 \\
    \vdots \\
    y^{DP}_N \\
    \end{array}
  \right)
  ~~~
  y_i^{DP}=\left(
    \begin{array}{c}
    y^{DP}_{i}(1) \\
    y^{DP}_{i}(2) \\
    \vdots \\
    y^{DP}_{i}(m^{DP}_i) \\
    \end{array}
  \right)
\end{align}
\end{small}
$y^t$ blocks for other types are similarly derived.

\subsubsection{Variables corresponding to discounts/surpluses of combined bids}
\item In the proposed model, while the standard bids are characterized by price per unit (PPU) bid prices and cleared based on MCP values represented by the variables $\varphi$, the combined bids are characterized by the package price -- in other words total maximal/minimal payoffs (regarding supply and demand respectively).
    While the SW in the case of the standard bids originates and may be calculated from the difference between market clearing MCPs (denoted by $\phi$), we assume that in the case of combined bids the maximal/minimal payoffs are subject to discount and surplus, which do also contribute to the total social welfare.
$W^D_{ij}$ and $W^S_{ij}$ denote the payoff discount of the combined demand bid $j$ submitted in node $i$ and the payoff surplus of the combined supply bid $j$ submitted in node $i$ respectively.

\begin{small}
\begin{align}
W=\left(
    \begin{array}{c}
      W^D \\
      W^S
    \end{array}
  \right)
  ~~~
  W^D=\left(
    \begin{array}{c}
      W^D_1 \\
      \vdots \\
      W^D_N
    \end{array}
  \right)
  ~~~
  W^D_i=\left(
    \begin{array}{c}
      W^D_{i1} \\
      \vdots \\
      W^D_{im^{DC}_i}
    \end{array}
  \right)
\end{align}
\end{small}
$W^S$ may be similarly derived.
$W^D \in \RR_+^{m^{DC}}$, $W^S \in \RR_+^{m^{SC}}$.

Any $W^D_{ij}$ or $W^S_{ij}$ may be greater than zero only, if the corresponding combined bid is accepted.

\end{itemize}

\subsubsection{The full state vector}

The full state vector of the model may be derived as
\begin{small}
\begin{equation}
x=\left(
    \begin{array}{c}
      v \\
      P \\
      Rp \\
      Rn \\
      \varphi \\
      y \\
      W \\
    \end{array}
  \right)
\end{equation}
\end{small}
$x \in \RR^{n(4T+1)+3NT+ m + m^{DC} +m^{SC}}$

\subsection{Cost model of the generating units}

The total cost of operation of the $j$th unit in node $i$, denoted by $C^G_{ij}$ is assumed to be linear and may be derived as
\begin{align}\label{generation_cost_1}
C^G_{ij}=\sum_t \beta_{ij}P_{ij}(t) + \alpha_{ij}v_{ij}(O)
\end{align}
where the first term describes the variable cost of production, depending on the output level of the unit, and the second term describes the start-up cost. The start up cost is considered in our framework for the whole modelled period. If the unit is e.g. turned off and on again in the analyzed time frame, it is considered as a warm start-up with negligible cost. However, based on the introduced variables $v_{ij}(t)$
the warm start-up costs may be taken into account as well, if necessary.
More complex and detailed formulations of start-up costs may be easily considered, following the methodology of the description of these costs in unit commitment approaches \cite{carrion2006computationally,viana2013new}.

We denote the total generation cost with $C^G$.
\begin{align}\label{generation_cost_total}
C^G=\sum_{ij} C^G_{ij}
\end{align}

\subsection{Auxiliary variables of the model}

We introduce a set of auxiliary variables, which will be used in the formulation of constraints and the objective functions. these variables do depend on the previously introduced primary model variables and on parameters.

\subsubsection{Demand side}

\paragraph{Power type variables}
We assume that the matrix $B^{DP}_i ~ \in \RR^{T \times m^{DP}_i}$ holds the bid quantities of the standard demand power bids corresponding to node $i$, given as a model parameter.
In this matrix, each row corresponds to a time period and each column corresponds to a bid. $B^{DP}_i(t,k)$ corresponds to the amount of the $k$-th demand power bid in node $i$ regarding time period $t$. For conventional bids, which regard only one period, only one element of the corresponding column is nonzero, while for block bids, multiple elements may be nonzero.

Similarly, $B^{DCP}_i~ \in \RR^{T \times m^{DC}_i}$ denotes the bid quantities corresponding to the power components of the combined demand bids submitted
in node $i$.

The total power demand at node $i$ in time period $t$, denoted by $D^P_i(t)$ may be derived as

\begin{small}
\begin{equation}\label{DP_i_t}
  D^P_i(t)=B^{DP}_i(t,.)y^{DP}_i~~+~~B^{DCP}_i(t,.)y^{DC}_i
\end{equation}
\end{small}
where $M(t,.)$ denotes the $t$-th row of the matrix $M$.

The total net power demand in time period $t$, denoted by $D^P(t)$ may be derived as

\begin{small}
\begin{equation}\label{DP_t}
  D^P(t)=\sum_i D^P_i(t)
\end{equation}
\end{small}

The total positive and negative reserve demands at node $i$ at time $t$, denoted by $D^{Rp}_i$ and $D^{Rn}_i$ respectively may be derived similarly
via the matrices $B^{DRp}_i$, $B^{DRn}_i$, $B^{DCRp}_i$ and $B^{DCRn}_i$ and the variables $y^{DRp}_i$, $y^{DRn}_i$, $y^{DC}_i$, $y^{DC}_i$.

\begin{small}
\begin{align}\label{DR_i_t}
&  D^{Rp}_i(t)=B^{DRp}_i(t,.)y^{DRp}_i ~~+~~B^{DCRp}_i(t,.)y^{DC}_i  \nonumber \\
&  D^{Rn}_i(t)=B^{DRn}_i(t,.)y^{DRn}_i ~~+~~B^{DCRn}_i(t,.)y^{DC}_i
\end{align}
\end{small}

The total net reserve demands for time period $t$ are as

\begin{small}
\begin{align}\label{DR_t}
  D^{Rp}(t)=\sum_i D^{Rp}_i(t)~~~~  D^{Rn}(t)=\sum_i D^{Rn}_i(t)
\end{align}
\end{small}

\paragraph{Income type variables}
We may decompose the total income ($I$) by nodes ($I_i$).

\begin{small}
\begin{align}\label{total income}
  I=\sum_i I_i
\end{align}
\end{small}

we may further decompose $I_i$ according to the various bid types
\begin{small}
\begin{align}\label{nodal income}
  I_i=I^{DP}_i + I^{DRp}_i + I^{DRn}_i +I^{DC}_i
\end{align}

\begin{align}\label{nodal income_DP}
  I^{DP}_i=\left(\varphi^P_i\right)^T B^{DP}_i y^{DP}_i
\end{align}
\end{small}
and similarly,
\begin{small}
\begin{align}\label{nodal income_DRp}
 & I^{DRp}_i=\left(\varphi^{Rp}_i\right)^T B^{DRp}_i y^{DRp}_i \\
 & I^{DRn}_i=\left(\varphi^{Rn}_i\right)^T B^{DRn}_i y^{DRn}_i
\end{align}
\end{small}
while for combined bids
\begin{small}
\begin{align}\label{nodal income_DC}
  I^{DC}_i= \left(y^{DC}_i\right)^T \left( \left(\Theta^{DC}_i\right)^T + W^D_i \right)
\end{align}
\end{small}

\subsubsection{Supply side}
In contrast to the demand side, the supply side of the model is composed not only from the FQ bids, but also from the FP bids. We use the upper indices $FQ$ and $FP$ to distinguish between these two types of bids.

\paragraph{Power type variables}
Similarly to the demand case, the matrix $B^{SP}_i ~ \in \RR^{T \times m^{SP}_i}$ describe the bid quantities of the standard supply power bids corresponding to node $i$.
$B^{SP}_i(j,k)$ corresponds to the amount of the $k$-th supply power bid in node $i$ regarding time period $j$.

Similarly, $B^{CSP}_i~ \in \RR^{T \times m^{SC}_i}$ denotes the bid quantities corresponding to the power components of the combined supply bids submitted
in node $i$.

The total power supply by FQ bids at node $i$ in time period $t$, denoted by $S^{PFQ}_i(t)$ may be derived as
\begin{small}
\begin{equation}\label{SP_FQ_i_t}
  S^{PFQ}_i(t)=B^{SP}_i(t,.)y^{SP}_i~~+~~B^{CSP}_i(t,.)y^{SC}_i
\end{equation}
\end{small}
The total net power supplied by FQ bids in time period $t$, denoted by $S^{PFQ}(t)$ is simply
\begin{small}
\begin{equation}\label{SP_FQ_t}
  S^{PFQ}(t)=\sum_i S^{PFQ}_i(t)
\end{equation}
\end{small}

The total positive and negative reserves supplied by FQ bids at node $i$ at time $t$, denoted by $S^{RpFQ}_i$ and $S^{RnFQ}_i$ respectively may be derived similarly
via the matrices $B^{SRp}_i$, $B^{SRn}_i$, $B^{SCRp}_i$ and $B^{SCRn}_i$ and the variables $y^{SRp}_i$, $y^{SRn}_i$, $y^{SC}_i$, $y^{SC}_i$.
\begin{small}
\begin{align}\label{SR_FQ_i_t}
&  S^{RpFQ}_i(t)=B^{SRp}_i(t,.)y^{SRp}_i ~~+~~B^{SCRp}_i(t,.)y^{SC}_i  \nonumber \\
&  S^{RnFQ}_i(t)=B^{SRn}_i(t,.)y^{SRn}_i ~~+~~B^{SCRn}_i(t,.)y^{SC}_i
\end{align}
\end{small}
The net balances are
\begin{small}
\begin{align}\label{SR_FQ_t}
 S^{RpFQ}(t)=\sum_i  S^{RpFQ}_i(t)~~~ S^{RnFQ}(t)=\sum_i  S^{RnFQ}_i(t)
\end{align}
\end{small}

\paragraph{Power type variables corresponding to FP bids}
For the description of the constraints corresponding to the sum of allocated power and reserve we will need the following variables.

Regarding FP bids, the $i$-th element of the following vector $\overline{S}^{PFP}(t)$ (denoted by $\overline{S}^{PFP}_i(t)$) describes the maximal power supplied by FP bids at node $i$ at time $t$.
\begin{small}
\begin{equation}\label{SP_FP_max_i_t}
  \overline{S}^{PFP}(t)=\overline{P}\left(
                           \begin{array}{c}
                             v_{11}(t) \\
                             \vdots \\
                             v_{1n_1}(t) \\
                             v_{21}(t) \\
                             \vdots \\
                             v_{Nn_N}(t) \\
                           \end{array}
                         \right)
\end{equation}
\end{small}
where $\overline{P}$ is a matrix holding the maximal production values of units, where the rows are corresponding to nodes.
$\overline{P}(i,k)\neq 0$ if and only if unit $k$ is located in node $i$.

Similarly, regarding minimal power output of FP bids
\begin{small}
\begin{equation}\label{SP_FP_min_i_t}
  \underline{S}^{PFP}(t)=\underline{P}\left(
                           \begin{array}{c}
                             v_{11}(t) \\
                             \vdots \\
                             v_{1n_1}(t) \\
                             v_{21}(t) \\
                             \vdots \\
                             v_{Nn_N}(t) \\
                           \end{array}
                         \right)
\end{equation}
\end{small}
where $\underline{P}$ is a matrix holding the minimal production values of units, where the rows are corresponding to nodes.
$\underline{P}(i,k)$ may be $\neq 0$ only if unit $k$ is located in node $i$.


Regarding the potential reserve production by FP bids, the $i$-th element of the following vector $\overline{S}^{RpFP}(t)$ (denoted by $\overline{S}^{RpFP}_i(t)$) describes the maximal positive reserve supplied by FP bids at node $i$ at time $t$.
\begin{small}
\begin{equation}\label{SRp_FP_max_i_t}
  \overline{S}^{RpFP}(t)=\overline{Rp}\left(
                           \begin{array}{c}
                             v_{11}(t) \\
                             \vdots \\
                             v_{1n_1}(t) \\
                             v_{21}(t) \\
                             \vdots \\
                             v_{Nn_N}(t) \\
                           \end{array}
                         \right)
\end{equation}
\end{small}
where $\overline{Rp}$ is a matrix holding the maximal positive reserve capacity values of units, where the rows are corresponding to nodes.
These values correspond to the theoretical maximum of positive reserve production, corresponding to technological constraints of the unit (load gradient constraints). The actual maximum of positive reserve may be lower compared to this value, e.g. if the plant is operating at maximum capacity, the actual available positive reserve is 0.

$\overline{Rp}(i,k)$ may be $\neq 0$ if and only if unit $k$ is located in node $i$.

The vector $\overline{S}^{RnFP}(t)$ corresponding to negative reserves may be derived similarly, via the matrix $\overline{Rn}$.


\paragraph{Cost type variables}
Similarly to the case of income, the total cost ($C$) is also decomposed by nodes ($C_i$).
Furthermore the payoff discounts ($W^D$) and payoff surpluses ($W^S$) are also considered as costs.
The reason for this is that in equations (\ref{nodal income_DC}) and (\ref{nodal cost}) the combined bids are accounted for with their nominal bid value, but thanks to payoff discounts and payoff surpluses ($W$) the incomes will be less and the costs will be higher (if the respective element of $W$ is greater than zero).
$W$ represents the contribution of these bids to the total social welfare.
\begin{small}
\begin{align}\label{total cost}
  C=\sum_i C_i + \sum_i W^{D}_i + \sum_i W^{S}_i
\end{align}
\end{small}

we may further decompose $C_i$ according to the various bid types
\begin{small}
\begin{align}\label{nodal cost}
  C_i=C^{SP}_i + C^{SRp}_i + C^{SRn}_i +C^{SC}_i + K^{FP}_i
\end{align}
\end{small}
Here the difference compared to Eq. (\ref{nodal income}) is the last term describing the payoff of FP bids, which is considered as a cost from the point of view of the auctioneer.
\begin{small}
\begin{align}\label{nodal cost_DP}
  C^{SP}_i=\left(\varphi^P_i\right)^T B^{SP}_i y^{SP}_i
\end{align}
\end{small}
and similarly,
\begin{small}
\begin{align}\label{nodal cost_DRp}
 & C^{SRp}_i=\left(\varphi^{Rp}_i\right)^T B^{SRp}_i y^{SRp}_i \\
 & C^{SRn}_i=\left(\varphi^{Rn}_i\right)^T B^{SRn}_i y^{SRn}_i
\end{align}
\end{small}
while for combined bids
\begin{small}
\begin{align}\label{nodal cost_DC}
  C^{SC}_i= \left(y^{SC}_i\right)^T \left( \left(\Theta^{SC}_i\right)^T + W^S_i \right)
\end{align}
\end{small}

\paragraph{Cost type variables corresponding to FP bids}
We decompose the nodal FP bid costs (payoffs) to individual costs of generating units as
\begin{small}
\begin{align}\label{nodal cost_FP}
  K^{FP}_i= \sum_j K^{FP}_{ij}~~~j \in \{1,...,n_i\}~~~K^{FP}=\sum_i K^{FP}_i
\end{align}
\end{small}
$K^{FP}_{ij}$ may be determined based on the actual allocated power and reserve production, and the relevant MCPs.
\begin{small}
\begin{align}\label{nodal cost_FP_individual}
  K^{FP}_{ij}=\sum_t P_{ij}(t)\varphi^P_i(t) + Rp_{ij}(t)\varphi^{Rp}_i(t) + Rn_{ij}(t)\varphi^{Rn}_i(t)
\end{align}
\end{small}


\subsection{Constraints}
In the following subsection the constraints of the model are summarized.

\subsubsection{Constraints corresponding to the range of variables}
First, we have to define constraints describing that power and reserves may be allocated only to active units, considering maximal and minimal output levels
\begin{small}
\begin{align}
& P_{ij}(t) + Rp_{ij}(t) \leq \overline{P}(i,j)v_{ij}(t)~~~\forall ~i,j,t \\
& P_{ij}(t) - Rn_{ij}(t) \geq \underline{P}(i,j)v_{ij}(t)~~~\forall ~i,j,t
\end{align}
\end{small}
Second, for acceptance indicators we have
\begin{small}
\begin{equation}\label{acceptance_indicators_range}
  0 \leq y \leq 1
\end{equation}
\end{small}
in addition, for block bids the corresponding $y$ values are binary.

\subsubsection{Bid acceptance constraints}
\label{Bid acceptance constraints}

\paragraph{1-hour bids}
In the case of 1-hour standard bids, the acceptance constraints are very simple. In the case of demand bids, they describe that the corresponding indicator variable $y^b_{ij} \geq 0$  if and only if the difference of the bid price and the relevant nodal price is nonnegative.

The matrix $\Theta^{DP}_i ~ \in \RR^{T \times m^{DP}_i}$ holds the bid PPUs of the standard power demand bids corresponding to node $i$.
In this matrix each column corresponds to a bid. $\Theta^{DP}_i(t,k)$ corresponds to the price of the $k$-th bid in node $i$, regarding time period $t$. For a conventional standard 1-hour bid, only one element in the corresponding column is nonzero, and its position is the same as of the nonzero element in the corresponding column in $B^{DP}_i$ (the matrix holding the bid quantities).
 $\Theta^{DRp}_i$ and $\Theta^{DRn}_i$ correspond to the prices of positive and negative standard demand reserve bids of node $i$.

In the case of 1-hour demand power bids we have the following rules
\begin{small}
\begin{align}
& y^{DP}_{ik}>0~~\rightarrow~~ \varphi^P_i(t_{rel}) \leq  \Theta^{DP}_i(t_{rel},k)~~~\forall k,i \nonumber \\
& y^{DP}_{ik}<1~~\rightarrow~~\varphi^P_i(t_{rel}) \geq \Theta^{DP}_i(t_{rel},k)~~~\forall k,i \label{bid_acc_rule_1_HR}
\end{align}
\end{small}
where $t_{rel}$ corresponds to the (relevant) time period, where the power demand corresponding quantity to $y^{DP}_{ik}$ is nonzero, which equals to the index of the nonzero element in the column vector $B^{DP}_i(.,k)$ of the matrix $B^{DP}_i$.

Similarly, in the case of 1-hour supply power bids

\begin{small}
\begin{align}
& y^{SP}_{ik}>0~~\rightarrow~~ \varphi^P_i(t_{rel}) \geq  \Theta^{SP}_i(t_{rel},k)~~~\forall k,i  \nonumber \\
& y^{SP}_{ik}<1~~\rightarrow~~\varphi^P_i(t_{rel}) \leq \Theta^{SP}_i(t_{rel},k)~~~\forall k,i
\end{align}
\end{small}

For the 1-hour positive/negative reserve demand/supply bids similar constraints may be derived \emph{mutatis mutandis}.

\paragraph{Block bids}
In the case of multiple-hour standard bids (block bids) we have to first define the SW value (denoted by $\Psi$) of the bid. In the case of demand bids, if the $j$-th bid of node $i$ is a block bid we have

\begin{small}
\begin{align}
&\Psi^{DP}_{ij}=\sum_t \Psi^{DP}_{ij}(t) \nonumber \\
&\Psi^{DP}_{ij}(t) = \left( B^{DP}_i(t,j)\cdot(\theta^{DP}_i(t,j)-\varphi^{P}_i(t))\right)
\end{align}
\end{small}
The corresponding constraint describes that the block bid is accepted if and only if its SW is positive.

\begin{small}
\begin{equation}\label{block_bid_acc_rule}
  y^{DP}_{ij}=1 ~~\Leftrightarrow~~\Psi^{DP}_{ij} > 0
\end{equation}
\end{small}
Again, for the positive/negative reserve demand/supply block bids (if such bids are present), similar constraints may be derived \emph{mutatis mutandis}.

\paragraph{Other bids} For other (combined and FP) bids we do not have explicit acceptance constraints, these bids are accepted or rejected by the clearing algorithm in order to maximize the total SW.

\subsubsection{Constraints corresponding to power and reserve balances}
\label{subsection_power_and_res_balances}
\paragraph{Global balances}
First, we have the global power balance equation as

\begin{small}
\begin{equation}\label{global_power_balance_1}
  D^P(t) - S^{PFQ}(t)  = P(t) =\sum_i P_i(t) ~~~~\forall t~.
\end{equation}
\end{small}

Regarding reserves, the total positive and negative reserve deficit by FQ bids must not exceed the potential maximal positive and negative reserve production by FP bids.

\begin{small}
\begin{align}
&  D^{Rp}(t) - S^{RpFQ}(t)  \leq \sum \overline{S}^{RpFP}(t)~~~~\forall t \nonumber \\
&  D^{Rn}(t) - S^{RnFQ}(t)  \leq \sum \overline{S}^{RnFP}(t)~~~~\forall t \label{global_reserve_balance}
\end{align}
\end{small}

\paragraph{Global combined balances}
In addition, since maximal power and nonzero positive reserve can not be allocated to any block in the same time, the sum of the net power deficit from FQ bids and the net positive reserve deficit from FQ bids must not exceed the maximal power amount which can be produced by the FP bids.

\begin{small}
\begin{equation}\label{global_power_and_pos_reserve_balance}
  D^P(t) - S^{PFQ}(t) + D^{Rp}(t) - S^{RpFQ}(t) \leq \sum \overline{S}^{PFT}(t)~~~~\forall t
\end{equation}
\end{small}

Similarly, the sum of the net power deficit and the net negative reserve deficit from FQ bids must be greater than the minimal amount which can be produced by the FP bids.

\begin{small}
\begin{equation}\label{global_power_and_neg_reserve_balance}
  D^P(t) - S^{PFQ}(t) - (D^{Rn}(t) - S^{RnFQ}(t)) \geq \sum \underline{S}^{PFT}(t)
\end{equation}
\end{small}

%
%
\subsubsection{Network constraints}
\label{subsection_network_constraints}

\paragraph*{Nominal case}
We assume that the constraints corresponding to the transmission network connecting the nodes are linear (consider e.g. a DC load flow approach), thus may be written in the form
\begin{small}
\begin{equation}\label{net_constr_1}
  A_{net}q(t) \leq b_{net}
\end{equation}
\end{small}
$A_{net} \in \RR^{N \times N}$ may be calculated as
\begin{small}
\begin{equation}\label{net_constr_2}
  A_{net}=E^DF^T E^{+}~~~,
\end{equation}
\end{small}
where $F\in \RR^{N \times K}$ is the node-branch incidence matrix of the network ($K$ denotes the number of lines, while $N$ is the number of nodes). $E \in \RR^{N \times N}$ denotes the susceptance matrix whose elements are $E_{kl}=-Y_{kl}$ for the off-diagonal terms and $$E_{kk}=-\sum_{l\neq k }^{} E_{kl}$$ (the column sum of off-diagonals) for diagonal elements. $Y_{kl}$ denotes the admittance of the line between nodes $k$ and $l$.
$E^{+}$ is the Moore-Penrose pseudoinverse of $E$, and $E^D$ is a diagonal matrix with $E^D_{kk}=Y_{ij}$.
The above formulation may be derived from the phase-angle approach described in \cite{Oren:1995} via the expression of the phase-angle vector as described in \cite{csercsik2017efficiency}. For further information on DC load flow models, see \cite{Oren:1995} and \cite{Contreras_thesis}.


%

The vector $b_{net}$ corresponds to the maximal power flow values of the lines. $q(t)\in \RR^N$ is the nominal power injection vector resulting from the market clearing. Its elements are corresponding to the power imbalances (= physical power injections) in each node. The $i$th element of $q(t)$, denoted by $q_i(t)$ corresponding to the power injection in node $i$ may be written as
\begin{small}
\begin{equation}\label{q}
  q_i(t)=S^{PFQ}_i(t) + P_i(t) - D^P_i(t)
\end{equation}
\end{small}
where
\begin{small}
$$
P_i(t)=\sum_jP_{ij}(t)
$$
\end{small}
Furthermore, according to the assumption regarding the lossless property of the network, which is usual in DC load flow models, we have
\begin{small}
\begin{equation}\label{power_injection_balance}
\sum_i q_i(t)=0 ~~~~\forall t
\end{equation}
\end{small}

\paragraph*{Perturbed case}
We require that the network constraints must hold also in the case when the allocated reserves are activated. If (positive) reserves are activated in a node, e.g. because of an unpredicted increase in the demand, the activation of the reserve has no consequences for the network. However it is possible that the cause of reserve activation is in another node.
We may view this scenario as a perturbed power injection vector $\hat{q}(t)$, for which the network must be also stable
\begin{small}
\begin{equation}\label{net_constr_2}
  A_{net}\hat{q}(t) \leq b_{net}~~~~\forall \hat{q}(t) \forall t
\end{equation}
\end{small}
where
\begin{small}
\begin{equation}\label{q_hat}
  \hat{q}(t)=q(t)+\delta(t)
\end{equation}
\end{small}
where $\delta(t) \in \RR^N$ is the perturbation vector, describing reserve activation. We assume that reserves may be activated at only one node in the same time, but in this case all of the allocated reserves (described by the total reserve demand $D^{Rp}_i(t)/D^{Rn}_i(t)$) are activated.
Furthermore, as we have $\sum q_i(t)=0$, as described in equation \ref{power_injection_balance},
we assume that the activated reserve must appear in a different node of the network with opposite sign.
Formally, regarding the $i$-th element of the vector $\delta$
\begin{small}
\begin{equation}\label{delta_i}
 (!\exists~~ i)~~  \left(\delta_i(t) \in \{D^{Rp}_i(t),D^{Rn}_i(t) \} \right)~~\left(!\exists~~ j \neq i\right)~~\left(\delta_j(t)=-\delta_i(t)\right)
\end{equation}
\end{small}
where $!\exists~~ i$ stands for 'there exists a unique $i$'.

\subsubsection{Scheduling constraints}
\label{subsection_scheduling_constraints}

\paragraph{Constraints corresponding to the overall activity indicator $v_{ij}(O)$}
The following constraints describe that if the block is active in any of the time periods, $v_{ij}(O)=1$, and 0 otherwise.
\begin{small}
\begin{align}
&\sum v_{ij} \leq Tv_{ij}(O)~~~~\forall (i,j)~~(i \in \{1,..N\})(j \in \{1,...n_i\} \nonumber \\
&v_{ij}(O) \leq \sum v_{ij}~~~~\forall (i,j)~~(i \in \{1,..N\})(j \in \{1,...n_i\}
\end{align}
\end{small}
\paragraph{Constraints corresponding to minimal up and down times}
Based on the introduced $v$ variables, these constraints may be derived in the same way as in unit commitment approaches \cite{carrion2006computationally,viana2013new} if necessary.

\paragraph{Load gradient constraints}
Load gradient constraints may be also formulated similar to unit commitment approaches, considering the possible activation of the allocated reserves as well.
\begin{small}
\begin{align}
&(P_{ij}(t+1)+Rp_{ij}(t+1))-(P_{ij}(t)-Rn_{ij}(t))<RU_{ij}~~~\forall~t<T \nonumber \\
&(P_{ij}(t)+Rp_{ij}(t))-(P_{ij}(t+1)-Rn_{ij}(t+1))<RD_{ij}~~~\forall~t<T
\end{align}
\end{small}
where $RU_{ij}$ and $RD_{ij}$ are the ramp-up and ramp-down constraints of the $j$-th unit in node $i$ respectively.

\subsubsection{Income and cost constraints}

\paragraph{Total income constraint}
The total income ($I$) from the demand bids must be at least equal to the cost of supply bids ($C$).
\begin{small}
\begin{equation}\label{total income_cost_balance}
  C \leq I
\end{equation}
\end{small}

\paragraph{Minimum income constraints of generating blocks}
\begin{small}
\begin{equation}\label{amount_paid_to_blocks}
K^{FP}=\sum_i K^{FP}_i
\end{equation}
\end{small}
where $K^{FP}_i$ is detailed in equations (\ref{nodal cost_FP}) and (\ref{nodal cost_FP_individual}).

From the point of view of the generating blocks this amount is an income, which has to cover the costs of generation

\begin{small}
\begin{equation}\label{MIC}
C^G_{ij} \leq K^{FP}_{ij}~~~~\forall (i,j)~~(i \in \{1,..N\})(j \in \{1,...n_i\})
\end{equation}
\end{small}

\paragraph{Distribution of discounts and surpluses among combined bids}
As we will see the objective function of the clearing model will be the maximization of total SW. The SW contribution of the standard bids may be calculated from MCPs, bid PPUs, and bid amounts. The SW contribution of FP bids is considered as the difference between their payoff ($K^{FP}_{ij}$) and their generating cost ($C^G_{ij}$).
The variables $W^D$ and $W^S$ represent the payoff discount and payoff surplus assigned to the
submitted combined bids. These variables may be viewed as follows. If we collect all income from the accepted demand bids and pay all costs regarding the supply bids (including FP and combined bids as well), thanks to the model constraints there will be a nonnegative residual, which may be divided among the accepted combined bids as payoff discounts or surpluses. In the following we define how this residual is distributed among the combined bids.

First, for each combined bid we define an average bid PPU denoted by $\lambda$.
We average over each quantity of the submitted combined bid, namely power, positive and negative reserve amounts. $\lambda^{D}_{ij}$ corresponds to the average PPU of the $j$-th combined demand bid submitted in node $i$ (average is understood over power and two types of reserve).
\begin{small}
\begin{equation}\label{combined_bid_aPPU}
  \lambda^{D}_{ij}=\frac{\sum B^{DCP}_i(.,j) + \sum B^{DCRp}_i(.,j) + \sum B^{DCRn}_i(.,j)}{\theta^{DC}_i(j)}
\end{equation}
\end{small}
$\lambda^{S}_{ij}$ may be derived similarly.

based on the above values, let us denote the minimal PPU among combined demand bids by $\underline{\lambda}^{D}$, and the maximal PPU among combined supply bids with $\overline{\lambda}^{S}$.

We assign the following weights to combined bids
\begin{small}
\begin{align}
&  a^D_{ij}=\frac{\lambda^{D}_{ij}-\underline{\lambda}^{D}}{\sum B^{DCP}_i(.,j) + \sum B^{DCRp}_i(.,j) + \sum B^{DCRn}_i(.,j)} \nonumber \\
\nonumber \\
&  a^S_{ij}=\frac{\overline{\lambda}^{D}-\lambda^{S}_{ij}}{\sum B^{SCP}_i(.,j) + \sum B^{SCRp}_i(.,j) + \sum B^{SCRn}_i(.,j)} \label{DC_weights}
\end{align}
\end{small}
Finally, we assume that the discount/surplus of every combined bid is proportional to its weight, considering only the accepted bids.
\begin{small}
\begin{align}\label{discount_surplus_distribution}
&  \frac{W^D_{ij}}{\sum W}=\frac{a^D_{ij}y^{DC}_{ij}}{\sum_{kl} a^D_{kl}y^{DC}_{kl} + \sum_{kl} a^S_{kl}y^{SC}_{kl} } ~~~~\forall (i,j)~~(i \in \{1,..N\})(j \in \{1,...n_i\})\nonumber \\
\nonumber \\
&  \frac{W^S_{ij}}{\sum W}=\frac{a^S_{ij}y^{SC}_{ij}}{\sum_{kl} a^D_{kl}y^{DC}_{kl} + \sum_{kl} a^S_{kl}y^{SC}_{kl} } ~~~~\forall (i,j)~~(i \in \{1,..N\})(j \in \{1,...n_i\})
\end{align}
\end{small}
We do not have to distinguish the accepted bids in $W$, since if the corresponding bid is not accepted, the representative element of $W$ is zero.
This results in a quadratic constraint after rearrangement, e.g. in the case of the demand side
\begin{small}
\begin{equation}\label{discount_surplus_distribution_2}
  W^D_{ij}\left(\sum_{kl} a^D_{kl}y^{DC}_{kl} + \sum_{kl} a^S_{kl}y^{SC}_{kl}\right)=a^D_{ij}y^{DC}_{ij}\sum W
\end{equation}
\end{small}
\subsection{The objective function}

The objective function of the model is to maximize the total SW, denoted by $\Psi$ which can be written as
\begin{small}
\begin{align}
 \Psi=&\Psi^{DP} + \Psi^{SP} + \Psi^{DRp} + \Psi^{SRp} + \Psi^{DRn} + \Psi^{SRn} +  \nonumber \\
      & K^{FP}-C_G + \sum W \nonumber \\
&\nonumber \\
&  \Psi^{DP}_{i}(t) = \left( B^{DP}_i(t,.)\odot(\theta^{DP}_i(t,.)-\varphi^{P}_i(t))\right)y^{DP}_i \nonumber \\
&  \Psi^{SP}_{i}(t) = \left( B^{SP}_i(t,.)\odot(\varphi^{P}_i(t)-\theta^{SP}_i(t,.))\right)y^{SP}_i
\nonumber \\
&  \Psi^{DRp}_{i}(t) = \left( B^{DRp}_i(t,.)\odot(\theta^{DRp}_i(t,.)-\varphi^{Rp}_i(t))\right)y^{DRp}_i
\nonumber \\
&  \Psi^{SRp}_{i}(t) = \left( B^{SRp}_i(t,.)\odot(\varphi^{Rp}_i(t)-\theta^{SRp}_i(t,.))\right)y^{SRp}_i
\nonumber \\
&  \Psi^{DRn}_{i}(t) = \left( B^{DRn}_i(t,.)\odot(\theta^{DRn}_i(t,.)-\varphi^{Rn}_i(t))\right)y^{DRn}_i
\nonumber \\
&  \Psi^{SRn}_{i}(t) = \left( B^{SRn}_i(t,.)\odot(\varphi^{Rn}_i(t)-\theta^{SRn}_i(t,.))\right)y^{SRn}_i
\nonumber \\
\end{align}
\end{small}
where the notation $\odot$ stands for the element-wise multiplication, and the notation $\theta^{DP}_i(t,.)-\varphi^{P}_i(t)$ stands for a vector, resulting from the element-wise subtraction of the scalar $\varphi^{P}_i(t)$ from the vector $\theta^{DP}_i(t,.)$.
In this formulation, regarding $\Psi^{DP}_{i}(t)$ we consider the accepted hourly and block bids together.
The notation is similar in the case of $\Psi^{SP}_{i}(t)$.



\section{Discussion}
\label{Discussion}

\subsection{Computational aspects}

 The balances and constraints for power and reserves described in subsection \ref{subsection_power_and_res_balances} are linear in the variables and do not pose a serious computational obstacle. Network constraints described in subsection \ref{subsection_network_constraints}, scheduling constraints discussed in subsection \ref{subsection_scheduling_constraints} are also linear. In the following we focus on less straightforward constraints: On the one hand on acceptance constraints derived from logical expressions (implications), and on the other hand on constraints involving quadratic terms of the variables.

\subsubsection{Bid acceptance constraints}
It is well known that in a combinatorial optimization framework logical expressions may be formulated in the terminology of computational constraints (see e.g. \cite{bemporad1999control}).
Bid acceptance constraints for FQ bids may be formulated with the application of auxiliary binary variables and the so called bigM method.

Let us consider the constraints described by eq. (\ref{bid_acc_rule_1_HR}), with a shorter notation

\begin{align}
y>0~~\rightarrow~~ \varphi \leq  \Theta~~~y<0~~\rightarrow~~ \varphi \geq  \Theta
\end{align}

where $y$ is the bid acceptance indicator, $\varphi$ is the MCP and $\Theta$ is the bid price. The formulation is equivalent to

\begin{align}
\varphi >  \Theta ~~\rightarrow~~ y=1~~~\varphi >  \Theta ~~\rightarrow~~ y=0 \label{acceptance_implication}
\end{align}

The former part of \ref{acceptance_implication} may be formulated as

\begin{align}
\varphi-\overline{\varphi}z \leq \Theta \nonumber \\
-y_1-(1-z)\leq -1 \label{bigM}
\end{align}
where $z$ is an auxiliary binary variable and $\overline{\varphi}$ is the upper bound for the MCP (the bigM in other words).

Regarding the acceptance rule of block bids described by the condition (\ref{block_bid_acc_rule}), we see that the form is analogous with the first expression of (\ref{bid_acc_rule_1_HR}), in the sense that a
$$
f_1(x)>b_1 ~~\rightarrow ~~ f_2(x) \leq b_2
$$
type implication is formulated, which is equivalent to $f_1(x) \leq b_1$ and/or $f_2(x) \leq b_2$.
This formulation is handled by eq. (\ref{bigM}).

\subsubsection{Constraints corresponding to combined bids}

 Regarding eq. (\ref{discount_surplus_distribution_2}) describing the distribution of surpluses and discounts of combined bids, we can see that it is a quadratic expression, which holds products of a binary and a continuous variables ($y$ and $W$ respectively). If we assume upper and lower bounds for $W$ (denoted by $\overline{W}$ and $\underline{W}$ respectively, from which $\underline{W}$ is potentially 0), and define the auxiliary variable $\zeta=y W$ for each product of this type, we may linearize the expression $\zeta=y W$ as
\begin{small}
\begin{align}
&\zeta\leq \overline{W} y \nonumber \\
& \zeta \geq \underline{W} y \nonumber \\
& \zeta \leq W-\underline{W}(1-y) \nonumber \\
& \zeta \geq W-\overline{W}(1-y)  \label{comp_bid_acc_4}
\end{align}
\end{small}

As potentially the number of combined bids is low in the market, such a linear reformulation implies a relatively low number of additional auxiliary variables ($\zeta$-s), so it is generally advised.

\subsubsection{Constraints describing minimum income conditions}

Probably the most difficult elements of the proposed formulation are the
 minimum income conditions described in eq. (\ref{MIC}), which include the terms $K^{FP_i}$, the payoff of flexible production bids, described by equation (\ref{nodal cost_FP_individual}). These are composed of quadratic expressions holding the product of continuous variables: The MCP's for power and reserves ($\varphi^P_i,~\varphi^{Rp}_i,~\varphi^{Rn}_i$) and power/reserve production quantities ($P_{i,j},~Rp_{i,j},~Rn_{i,j}$) -- thus they are a critical point regarding computational issues. If such flexible-production units and constraints are present, the implied problem falls into the class of non-convex quadratically constrained quadratic (QCQP) programs.
The more recent advances on such problems are described in \cite{elloumi2019global}.
 Further papers discuss the possible approaches for this problem class, as exact quadratic convex reformulation \cite{billionnet2016exact} or piecewise linear and edge-concave relaxations \cite{misener2012global}.
Results corresponding to general non-convex mixed-integer nonlinear problems are surveyed in \cite{burer2012non}.

%
%

Despite the recent advances described in \cite{elloumi2019global}, QCQP is still considered as a hard problem class, for which large-scale implementations pose a significant challenge.
 Regarding this issue, let us however recite the consideration described in \cite{galiana2005scheduling}, namely that it is likely that steady progress in computing technologies could well curb the above difficulties within the coming years.

\subsection{Prospects for generalizations}

How the proposed model could be generalized for the procurement of multiple (i.e. secondary and tertiary) reserves simultaneously is a straightforward question. In the context of the described framework,
tertiary reserves would mean reserves with higher response time, but otherwise they are considered as the reserves discussed before (capacity allocation payment is assumed).
Such an extension is possible, however it would significantly increase the complexity of the model.

Naturally, in such a framework regarding one-hour and multiple hour single product bids the secondary (S) and tertiary (T) reserves must be distinguished, as well as the MCPs for which additional variables shall be defined. Regarding FP bids, instead the variables $Rp$ and $Rn$ similar variables as $RSp$, $RSn$, $RTp$ $RTn$ should be used corresponding to allocated amounts of secondary and tertiary reserves. Addition parameters (maximally allocated S and T type reserves must be also considered).

Additional constraints in this case have be included to describe the asymmetry of substitution relations -- e.g. the sum of allocated S and T type reserves for any hour must not exceed the maximal amount of T type reserve which may be allocated, and so on. Ramp limits of the units must be considered to formulate such constraints. The combined bids would be the least problematic elements in this framework. they only have to be extended with an amount regarding T type reserves, but every other aspects of them may stay the same.

Auxiliary variables of course must be updated/extended as well (e.g. the total net demand for S and T type reserves must be distinguished, etc.). Constraints corresponding to the range of variables must be updated, as the sum of allocated power, S and T type of positive reserves must not exceed the maximal production value $\overline{P}$ (and mutatis mutandis in the case of $\underline{P}$).
Balances must be updated as well. Inequalities \ref{global_reserve_balance} must be formulated distinctly for S and T type reserves, and global combined balances (eq. \ref{global_power_and_pos_reserve_balance}) also have to be updated.

Network constraints in this case must consider perturbed power injection vectors corresponding to the activation of tertiary reserves as well. Load gradient constraints must be formulated considering both types of reserve, and naturally the income and cost constraints must me modified as well to account for the now product type. In the objective function, the new terms corresponding to $T$ type reserve bids must be included.

\section{Conclusions}
The formulation of SW based simultaneous clearing methods for power and ancillary services is a complex task even in the case when the network constraints are neglected. In the current paper we proposed a market coupling approach of integrated power-reserve markets including innovative orders which could help the efficient bidding of generating units and by adding additional bidding alternatives make the market more flexible. In addition the proposed formulation also includes network constraints for the nominal (or undisturbed) case and also considers scenarios when the reserves are activated. The described approach results in a computationally hard, but likely not out-of reach problem.

\section{Acknowledgements}
This work has been supported by the Fund PD 123900 of the Hungarian National Research, Development and Innovation Office, and by the J\'{a}nos Bolyai Research Scholarship of the Hungarian Academy of Sciences.

 \bibliographystyle{plain}
\bibliography{energy_GT_1_abr.bib}

\newpage
\section*{Appendix A: Nomenclature}

\begin{table}[h!]
\begin{tabular}{|c|c|c|}
  \hline
  Parameter & interpretation & dimension \\
  \hline
  $T$ & Number of time periods & - \\ \hline
  $N$ & Number of nodes in the transmission network & - \\ \hline
  $K$ & Number of lines in the transmission network & - \\ \hline
  $n_i$ & \begin{tabular}{@{}c@{}} number of generating units submitting \\ flexible production (FP) bids in node $i$\end{tabular} & - \\ \hline
  $n$ & \begin{tabular}{@{}c@{}} Total number of generating units submitting \\ flexible production (FP) bids \end{tabular} & - \\ \hline
  $m^{DP}_i$ & the number of standard demand power bids at node $i$ & - \\ \hline
  $m^{DP}$ & the total number of standard demand power bids& - \\ \hline
  $m^{SP}_i$ & the number of standard supply power bids at node $i$. & - \\ \hline
  $m^{SP}$ & the total number of standard supply power bids  & - \\ \hline

  $m^{DRp}_i$ & the number of standard demand positive reserve bids at node $i$ & - \\ \hline
  $m^{DRp}$ & the total number of standard demand positive reserve bids & - \\ \hline
  $m^{DRn}_i$ & the number of standard demand negative reserve bids at node $i$ & - \\ \hline
  $m^{DRn}$ & the total number of standard demand negative reserve bids & - \\ \hline

  $m^{SRp}_i$ & the number of standard supply positive reserve bids at node $i$ & - \\ \hline
  $m^{SRp}$ & the total number of standard supply positive reserve bids & - \\ \hline
  $m^{SRn}_i$ & the number of standard supply negative reserve bids at node $i$ & - \\ \hline
  $m^{SRn}$ & the total number of standard supply negative reserve bids & - \\ \hline
  $m^{DC}_i$ & the number of combined demand bids at node $i$ & - \\ \hline
  $m^{DC}$ & the total number of combined demand bids& - \\ \hline
  $m^{SC}_i$ & the number of combined supply bids at node $i$ & - \\ \hline
  $m^{SC}$ & the total number of combined supply bids& - \\ \hline
  $m$ & the total number of fixed quantity (FQ) bids& - \\
  \hline
\end{tabular}

\end{table}

\newpage

\begin{table}[h!]
\begin{tabular}{|c|c|c|}
  \hline
  Parameter & interpretation & dimension \\
  \hline
  $B^{DP}_i$ &  \begin{tabular}{@{}c@{}} Matrix of bid quantities of standard \\ demand power bids corresponding to node $i$ \end{tabular}  & MW \\ \hline
  $B^{DRp}_i$ &  \begin{tabular}{@{}c@{}} Matrix of bid quantities of standard positive \\ reserve demand bids corresponding to node $i$ \end{tabular}  & MW \\ \hline
  $B^{DNp}_i$ &  \begin{tabular}{@{}c@{}} Matrix of bid quantities of standard negative \\ reserve demand bids corresponding to node $i$ \end{tabular}  & MW \\ \hline
  $B^{DCP}_i$ &  \begin{tabular}{@{}c@{}} Matrix of bid quantities corresponding to the power components \\ of the combined demand bids submitted in node $i$ \end{tabular}  & MW \\ \hline
  $B^{DCRp}_i$ &  \begin{tabular}{@{}c@{}} Matrix of bid quantities corresponding to the positive reserve \\ components of the combined demand bids submitted in node $i$ \end{tabular}  & MW \\ \hline
  $B^{DCRn}_i$ &  \begin{tabular}{@{}c@{}} Matrix of bid quantities corresponding to the negative reserve \\ components of the combined demand bids submitted in node $i$ \end{tabular}  & MW \\ \hline
  $B^{SP}_i$ &  \begin{tabular}{@{}c@{}} Matrix of bid quantities of standard \\ supply power bids corresponding to node $i$ \end{tabular}  & MW \\ \hline
  $B^{SRp}_i$ &  \begin{tabular}{@{}c@{}} Matrix of bid quantities of standard positive \\ reserve supply bids corresponding to node $i$ \end{tabular}  & MW \\ \hline
  $B^{SNp}_i$ &  \begin{tabular}{@{}c@{}} Matrix of bid quantities of standard negative \\ reserve supply bids corresponding to node $i$ \end{tabular}  & MW \\ \hline
  $B^{SCP}_i$ &  \begin{tabular}{@{}c@{}} Matrix of bid quantities corresponding to the power components \\ of the combined supply bids submitted in node $i$ \end{tabular}  & MW \\ \hline
  $B^{SCRp}_i$ &  \begin{tabular}{@{}c@{}} Matrix of bid quantities corresponding to the positive reserve \\ components of the combined supply bids submitted in node $i$ \end{tabular}  & MW \\ \hline
  $B^{SCRn}_i$ &  \begin{tabular}{@{}c@{}} Matrix of bid quantities corresponding to the negative reserve \\ components of the combined supply bids submitted in node $i$ \end{tabular}  & MW \\ \hline
  $\overline{P}$ &  Matrix of the maximal production values of units  & MW \\ \hline
  $\underline{P}$ &  Matrix of the minimal production values of units  & MW \\   \hline
  $\overline{Rp}$ &  \begin{tabular}{@{}c@{}} Matrix of the upper bounds for positive reserve \\ allocation values of units \end{tabular} & MW \\ \hline
  $\overline{Rn}$ &  \begin{tabular}{@{}c@{}} Matrix of the upper bounds for negative reserve \\ allocation values of units \end{tabular}  & MW \\   \hline
  $\alpha_{ij}$ & fixed cost of operation of unit $j$ of node $i$ & EUR \\ \hline
  $\beta_{ij}$ & variable cost of production of unit $j$ of node $i$ & EUR/MW \\ \hline
  $A_{net}$ &  matrix of transmission network constraints  & - \\ \hline
  $b_{net}$ &  vector of maximal power flow of network lines  & MW \\ \hline
  $RU_{ij}$ &  ramp-up limit of unit $j$ in node $i$   & MW \\ \hline
  $RD_{ij}$ &  ramp-down limit of unit $j$ in node $i$   & MW \\ \hline
  $E$ &  susceptance matrix   & $\Omega^{-1}$ \\ \hline
  $F$ & the node-branch incidence matrix of the network & - \\ \hline
  $Y_{k,l}$ & susceptance of the line between nodes $k$ and $l$  & $\Omega^{-1}$  \\
  \hline
\end{tabular}

\end{table}

\newpage

\begin{table}[h!]
\begin{tabular}{|c|c|c|}
  \hline
  Parameter & interpretation & dimension \\
  \hline
  $\Theta^{DP}_i$ & \begin{tabular}{@{}c@{}} bid PPU matrix of the standard power \\ demand bids corresponding to node $i$ \end{tabular} & EUR/MW \\ \hline
  $\Theta^{DRp}_i$ & \begin{tabular}{@{}c@{}} bid PPU matrix of the standard positive reserve   \\ demand bids corresponding to node $i$ \end{tabular}  & EUR/MW \\ \hline
  $\Theta^{DRn}_i$ & \begin{tabular}{@{}c@{}} bid PPU matrix of the standard negative reserve   \\ demand bids corresponding to node $i$ \end{tabular} & EUR/MW \\ \hline
  $\Theta^{SP}_i$ & \begin{tabular}{@{}c@{}} bid PPU matrix of the standard power \\ supply bids corresponding to node $i$ \end{tabular}  & - \\ \hline
  $\Theta^{SRp}_i$ & \begin{tabular}{@{}c@{}} bid PPU matrix of the standard positive reserve  \\ supply bids corresponding to node $i$ \end{tabular}  & EUR/MW \\ \hline
  $\Theta^{SRn}_i$ & \begin{tabular}{@{}c@{}} bid PPU matrix of the standard negative reserve  \\ supply bids corresponding to node $i$ \end{tabular}  & EUR/MW \\ \hline
  $\Theta^{DC}_i$ & \begin{tabular}{@{}c@{}} bid price (row) vector of the combined demand \\ bids corresponding to node $i$ \end{tabular}    & EUR \\ \hline
  $\Theta^{SC}_i$ & \begin{tabular}{@{}c@{}} bid price (row) vector of the combined supply \\ bids corresponding to node $i$ \end{tabular}    & EUR \\ \hline
  $\lambda^{D}_{ij}$ & \begin{tabular}{@{}c@{}} average PPU of the $j$th combined demand bid \\ submitted in  node $i$ \end{tabular} & EUR/MW \\ \hline
  $\lambda^{S}_{ij}$ & \begin{tabular}{@{}c@{}} average PPU of the $j$th combined suply bid \\ submitted in  node $i$ \end{tabular} & EUR/MW \\ \hline
  $\underline{\lambda}^{D}$ & minimal average PPU among combined demand bids & EUR/MW \\ \hline
  $\overline{\lambda}^{S}$ & maximal average PPU among combined demand bids & EUR/MW \\ \hline
  $a^D_{ij}$ & \begin{tabular}{@{}c@{}} weight of the $j$th combined demand bid \\ submitted in  node $i$ \end{tabular} & EUR \\ \hline
  $a^S_{ij}$ & \begin{tabular}{@{}c@{}} weight of the $j$th combined supply bid \\ submitted in  node $i$ \end{tabular} & EUR \\ \hline
\end{tabular}
\end{table}
\newpage
Variables

\begin{table}[h!]
\begin{tabular}{|c|c|c|}
  \hline
  Variable & interpretation & dimension \\ \hline
 $v_{ij}(t)$ & activity indicator of block $j$ of node $i$ at time $t$ & - \\ \hline
  $v_i(t)$ & vector of activity indicators of node $i$ at time $t$   & - \\ \hline
  $v(t)$ & vector of activity indicators at time $t$   & - \\ \hline
  $P_{ij}(t)$ &  power production value allocated to block $j$ of node $i$ at time $t$  & MW \\ \hline
  $P_{i}(t)$ &  total power production value of node $i$ at time $t$  & MW \\ \hline
  $P(t)$ &  total power production value at time $t$  & MW \\ \hline
  $Rp_{ij}(t)$ &  positive reserve value allocated to block $j$ of node $i$ at time $t$  & MW \\ \hline
  $Rp_{i}(t)$ &  total positive reserve value allocated node $i$ at time $t$  & MW \\ \hline
  $Rn_{ij}(t)$ &  negative reserve value allocated to block $j$ of node $i$ at time $t$  & MW \\ \hline
  $Rn_{i}(t)$ &  total negative reserve value of node $i$ at time $t$  & MW \\ \hline
  $\varphi_i^P(t)$ &  the market clearing price (MCP) of power at node $i$ at time $t$  & EUR/MW \\ \hline
  $\varphi_i^{Rp}(t)$ & MCP of positive reserve at node $i$ at time $t$   & EUR/MW \\ \hline
  $\varphi_i^{Rn}(t)$ & MCP of negative reserve at node $i$ at time $t$   & EUR/MW \\ \hline
  $y^b_{ij}$ &  \begin{tabular}{@{}c@{}} bid acceptance indicator of the $j$-th bid of type $b$ \\ submitted in node $i$ \end{tabular} & - \\ \hline
  $W^D_{ij}$ &  \begin{tabular}{@{}c@{}} Payoff discount of the $j$-th combined demand bid \\ submitted in node $i$ \end{tabular}  & EUR \\ \hline
  $W^S_{ij}$ &  \begin{tabular}{@{}c@{}} Payoff surplus of the $j$-th combined supply bid \\ submitted in node $i$ \end{tabular}  & EUR \\ \hline
  $\sum W$ &  Sum of payoff surpluses and discounts  & EUR \\ \hline
\end{tabular}
\end{table}

\newpage
Auxiliary variables
\begin{table}[h!]
\begin{tabular}{|c|c|c|}
  \hline
  Variable & interpretation & dimension \\ \hline
  $D^P_i(t)$ & Total power demand at node $i$ in time period $t$    & MW \\ \hline
  $D^P(t)$ &  Total net power demand in time period $t$ & MW \\ \hline
  $D^{Rp}_i(t)$ &  Total positive reserve demands at node $i$ at time $t$  & MW \\ \hline
  $D^{Rn}_i(t)$ &  Total negative reserve demands at node $i$ at time $t$  & MW \\ \hline
  $D^{Rp}(t)$ &  Total net positive reserve demand at time $t$  & MW \\ \hline
  $D^{Rn}(t)$ &  Total net negative reserve demand at time $t$  & MW \\ \hline
  $S^{PFQ}_i(t)$ &  Total power supply by FQ bids at node $i$ in time period $t$  & MW \\ \hline
  $S^{PFQ}(t)$ &  Total net power supply by FQ bids in time period $t$  & MW \\ \hline
  $S^{RpFQ}_i$ & \begin{tabular}{@{}c@{}} Total amount of positive reserves supplied by FQ bids \\ at node $i$ at time $t$ \end{tabular} & MW \\  \hline
  $S^{RnFQ}_i$ & \begin{tabular}{@{}c@{}} Total amount of negative reserves supplied by FQ bids \\ at node $i$ at time $t$ \end{tabular}  & MW \\  \hline
  $S^{RpFQ}$ & \begin{tabular}{@{}c@{}} Total net amount of positive reserves supplied \\ by FQ bids at time $t$ \end{tabular}   & MW \\  \hline
  $S^{RnFQ}$ & \begin{tabular}{@{}c@{}} Total net amount of negative reserves supplied \\ by FQ bids at time $t$ \end{tabular}   & MW \\  \hline
  $q(t)$ & power injection vector   & MW \\  \hline
  $\hat{q}(t)$ & perturbed power injection vector   & MW \\  \hline
  $\delta$ & perturbation vector, describing reserve activation   & - \\  \hline
  $I$ & Total income   & EUR \\  \hline
  $I_i$ & Income in node $i$   & EUR \\  \hline
  $I^{DP}_i$ & Income in node $i$ from power demand bids  & EUR \\  \hline
  $I^{DRp}_i$ & Income in node $i$ from positive reserve demand bids  & EUR \\  \hline
  $I^{DRn}_i$ & Income in node $i$ from negative reserve demand bids  & EUR \\  \hline
  $I^{DC}_i$ & Income in node $i$ from combined bids  & EUR \\  \hline
  $C$ & Total cost   & EUR \\  \hline
  $C_i$ & Cost in node $i$   & EUR \\  \hline
  $C^{DP}_i$ & Cost of power demand bids in node $i$  & EUR \\  \hline
  $C^{DRp}_i$ & Cost of positive reserve demand bids in node $i$  & EUR \\  \hline
  $C^{DRn}_i$ & Cost of negative reserve demand bids in node $i$  & EUR \\  \hline
  $C^{DC}_i$ & Cost of combined bids in node $i$   & EUR \\  \hline
  $K^{FP}_{ij}$ & Payoff of the $j$th generating unit in node $i$  & EUR \\  \hline
  $K^{FP}_i$ & Payoff of FP bids in node $i$  & EUR \\  \hline
  $K^{FP}$ & Total payoff of FP bids & EUR \\  \hline
  $C^{G}_{ij}$ & Generating cost of the $j$th generating unit in node $i$  & EUR \\  \hline
  $C^{G}_i$ & Generating cost of units in node $i$  & EUR \\  \hline
  $C^{G}$ & Total generating cost of units submitting FP bids & EUR \\  \hline
  $\Psi$ & total social welfare   & EUR \\  \hline
  $z$    & computational auxiliary variable for bid acceptance constraints & - \\ \hline
  $\zeta$ & comp. auxiliary variable for surplus distribution constraints & - \\ \hline
\end{tabular}

\end{table}
\newpage
The argument notation in the case of a variable represents the corresponding time period, e.g. $D^P_i(t)$ denotes the total power demand at node $i$ in time period $t$, while in the case of parameter vectors, it represents the corresponding element, e.g. $\Theta^{DC}_i(j)$ stands for the $j$th element of the vector
$\Theta^{DC}_i$. Two arguments are used in the case of matrices, e.g. $\overline{P}(i,j)$ stands for the element in the $i$th row and $j$th column of the matrix $\overline{P}$. $M(.,j)$ and $M(i,.)$ denote the $j$th column and the $i$th row of the matrix $M$ respectively.

\end{document}